\documentclass[manuscript]{aastex}

\usepackage{epsfig}
\usepackage{ulem}

\def\'#1{\ifx#1i{\accent"13\i}\else{\accent"13#1}\fi}

\newcommand{\kms}{{\rm ~km~s}^{-1}}
\newcommand{\ppcc}{{\rm ~cm}^{-3}}
\newcommand{\gpcc}{{\rm ~g~cm}^{-3}}
\newcommand{\Myr}{{\rm Myr}}
\newcommand{\Mmax} {M_{\rm max}}
\newcommand{\Mdens} {M_{\rm dens}}
\newcommand{\ndens} {n_{\rm dens}}
\newcommand{\Mc}{M_{\rm C}}
\newcommand{\Rc}{R_{\rm C}}
\newcommand{\Rinf}{R_{\rm inf}}
\newcommand{\Mrms}{{\cal M}_{\rm rms}}
\newcommand{\tSF}{t_{\rm SF}}
\newcommand{\Mstar}{M_*}
\newcommand{\Mi}{M_{\rm I}}
\newcommand{\NC}{N_{\rm C}}
\newcommand{\nsf} {n_{\rm SF}}

\newcommand{\rhoavg}{\langle{\rho}\rangle}
\newcommand{\navg}{\langle{n}\rangle}
\newcommand{\SFR}{{\rm SFR}}
\newcommand{\SFRavg}{\langle{\rm SFR}\rangle}

\newcommand{\SFEavg}{\langle{\rm SFE}\rangle}
\newcommand{\SFEmax}{{\rm SFE}_{\rm final}}

\newcommand{\tff}{{t_{\rm ff}}}
\newcommand{\Msun}{{\rm M}_{\odot}}
\newcommand{\VS}{V\'azquez-Semadeni}
\newcommand{\ZA}{Zamora-Avil\'es }

\newcommand{\ii}{\'\i}
\newcommand{\etal}{et al.}
\newcommand{\beq}{\begin{equation}}
\newcommand{\eeq}{\end{equation}}

\slugcomment{Submitted to ApJ}

\shorttitle{An Evolutionary Model for the SFR. II. Mass Dependence}
\shortauthors{Zamora-Avil\'es and V\'azquez-Semadeni}

\begin{document}


\title{An Evolutionary Model for Collapsing Molecular Clouds and Their
Star Formation Activity. II. Mass Dependence of the Star Formation Rate}


\author{Manuel Zamora-Avil\'es\altaffilmark{1} \& Enrique V\'azquez-Semadeni\altaffilmark{1}}


\altaffiltext{1}{Centro de Radioastronom\'\i a y Astrof\'\i sica,
Universidad Nacional Aut\'onoma de M\'exico, Apdo. Postal 3-72, Morelia,
Michoac\'an, 58089, M\'exico}


\begin{abstract}

We discuss the evolution, and dependence on cloud mass, of the star
formation rate (SFR) and efficiency (SFE) of star-forming molecular
clouds (MCs), within the scenario that clouds are undergoing global
collapse, and that the SFR is controlled by ionization feedback. We find 
that low-mass clouds ($\Mmax \lesssim 10^4~\Msun$) spend most of their
evolution at low SFRs, but end their lives with a mini-burst, reaching a
peak SFR $\sim 10^4 ~\Msun \Myr^{-1}$, although their time-averaged SFR is
only $\SFRavg \sim 
10^2 ~ \Msun \Myr^{-1}$. The
corresponding efficiencies are $\SFEmax \lesssim $60\% and $\SFEavg
\lesssim$1\%. For more massive clouds ($\Mmax \gtrsim 10^5 ~ \Msun$),
the SFR first increases and then reaches a plateau, because the clouds
are influenced by the stellar feedback since earlier in their
evolution. As a function of cloud mass, $\SFRavg$ and $\SFEavg$ are well
represented by the fits $\SFRavg \approx 100 (1+\Mmax/1.4 \times 10^5 ~
\Msun)^{1.68} ~ \Msun \Myr^{-1}$ and $\SFEavg \approx 0.03 (\Mmax/2.5
\times 10^5 ~ \Msun)^{0.33}$, respectively. Moreover, the SFR of our
model clouds follows closely the SFR-dense gas mass relation recently
found by Lada et al., during the epoch when their instantaneous SFEs are
comparable to those of the clouds considered by those authors.
Collectively, a Monte Carlo integration of the model-predicted $\SFR(M)$
over a Galactic GMC mass spectrum yields values for the total Galactic
SFR that are within half an order of magnitude from the relation
obtained by Gao \& Solomon.  Our results support the scenario that
star-forming MCs may be in global gravitational collapse, and that the
low observed values of the SFR and SFE are a result of the
interruption of each SF episode, caused primarily by the ionizing
feedback from massive stars.

\end{abstract}


\section{Introduction} \label{sec:intro}

The regulation of the star formation rate (SFR) in molecular clouds
(MCs) has been a key problem in astrophysics for over half a century,
ever since \citet{Schmidt59} noticed that the SFR in clouds exhibited a
power-law dependence on the gas number density $n$.  A crucial aspect 
of the SFR was noticed by
\citet{ZP74}, who pointed out that the observed Galactic SFR is at least
one order of magnitude lower than that expected if the clouds were
forming stars at the ``free-fall rate'', given by the ratio of the total
molecular gas mass in the Galaxy to the typical free-fall time of this
gas. Indeed, current estimates of the total molecular gas mass and
density \citep[$M_{\rm mol} \sim 10^9~\Msun$, $n \sim 100
\ppcc$; e.g.,][] {Ferriere01} imply a free-fall SFR $\sim 200~\Msun$
yr$^{-1}$, while the observed SFR is roughly 100 times smaller
\citep[e.g.,][]{ChP11}. Thus, it was concluded that MCs could not be in
free-fall, contrary to the then recent suggestion of \citet{GK74}, and
that the nonthermal linewidths observed in the clouds were produced by
small-scale turbulence instead \citep{ZE74}.

Since then, MCs have been assumed to be supported by a number of
physical agents, such as magnetic fields \citep[e.g.,][] {SAL87,
Mousch91} or turbulence \citep[e.g.,][] {VS+00, VS+03, MK04, ES04, BP+07,
MO07}. In both scenarios, the necessary low SFR was attained because a
small fraction of the mass managed to escape the support. This fraction
was mediated by ambipolar diffusion in the first case, and by local
turbulent compressions which induced small-scale, low-mass collapses in
the second. In the last decade, a number of models for the turbulent
regulation of star formation (SF) have been constructed within the
scenario of clouds in which both the global support and the local
collapses are induced by turbulence \citep{KM05, PN11, HC11}. These
models are based on the premise that the high-density tail of the
density probability density function (PDF), which takes a lognormal form
for supersonic isothermal turbulence \citep{VS94, Padoan+97, PV98}, is
responsible for the instantaneous SFR, which is given by this mass,
divided by a characteristic timescale. The models differ in how the
threshold density for defining the ``high-density'' gas and the
characteristic timescale.  A thorough discussion of these models has
been recently provided by
\citet{FK12}. 

However, recent evidence from both observations and numerical
simulations has suggested that star-forming MCs may be in gravitational
collapse after all. Comparing numerical simulations of a variety of
turbulent and free-falling regimes to the observed kinematics of the
clump NGC 2264-C, \citet{PHA07} showed that the best fit was provided by
simulations in which infall dominates over turbulence by a large margin
(95\% of the kinetic energy). Comparing the morphology of the Orion A
cloud to that of simulations of gravitational collapse of a nearly
elliptical sheet of gas, \citet{HB07} suggested that the entire Orion A
cloud may be in gravitational collapse. Also, infall has been observed
at multiple scales in the high-mass star forming region G20.08-0.14
\citep{Galvan+09} and from filamentary regions onto clumps, as well as
onto the filaments \citep{Schneider+10, Kirk+13}. On the numerical side,
simulations of cold, dense cloud formation including self-gravity
\citep{VS+07, VS+09, VS+10, VS+11, HH08} have shown that the clouds engage in
gravitational collapse shortly after they collect enough mass to be
Jeans-unstable, and long before any star formation begins to occur
within them. 
Moreover, the nonlinear density fluctuations produced by the
turbulence in the cloud have shorter free-fall timescales than the cloud
at large, and therefore complete their collapses before the cloud does.
Thus, \citet{VS+09} suggested that MCs are in a state of ``hierarchical
gravitational collapse'', where the {\it local}, small-scale collapses
of dense cores are occurring within the {\it global}, large-scale
collapse of the cloud.

In a previous paper \citep[][hereafter Paper I]{Zamora+12}, we presented
an analytical model for the evolution of the SFR in the context of
gravitationally collapsing clouds. This model was based on the same
prescription for computing the SFR as that
used in the models mentioned above; i.e., an integration of the
density PDF above a certain threshold density to obtain the mass
responsible for the ``instantaneous'' SFR. The threshold density was
obtained by calibrating the evolution of the SFR with the numerical
simulations of \citet{VS+10}, and the timescale was chosen as the
free-fall time at the threshold density. However, the distinctive
feature of the model was that the cloud, assumed to have a sheet-like
geometry, was considered to be undergoing free-fall gravitational
collapse, causing its mean density to increase. Therefore, the density
PDF was considered to continuously shift to higher densities, and thus
the star-forming mass continuously increased in time, implying a
systematic increase of the SFR. The controlling parameter of the model
was found to be the total system mass, and the model successfully
described the evolutionary sequence for $\sim 10^5~\Msun$ GMCs reported
by \citet{Kawamura+09}, the stellar age histograms for clouds of mass
$\sim 2000~\Msun$, as reported by \citet{PS00, PS02}, and the locus
of clouds of this same mass in the Kennicutt-Schmidt diagram, as
reported by \citet{Evans+09}. 

Since the main controlling parameter of the model from Paper I was the
cloud's mass, in this paper we now examine the predictions of the model
for the dependence of the SFR and the star formation efficiency (SFE) with
the mass of the cloud,
and from there examine the prediction of the model for the SFR-mass
relation first proposed by \citet[][hereafter, the GS
relation]{Gao+04}. The plan of the paper is as follows: in Sec.\
\ref{sec:model}, we present a brief summary of the model, as well as its
application to the present study. In Sec.\ \ref{sec:predictions} we present the
results for the dependence of the SFR and the SFE with cloud mass, and
compare the model with the observational GS relation. Then, in Sec.\
\ref{sec:discussion} we discuss some implications and limitations of our results. 
Finally, in Sec.\ \ref{sec:sum} we present a summary.

\section{The Model} \label{sec:model}

{\bf Our model, first presented in Paper I, follows the evolution of the
gas mass that initially constitutes a cold atomic cloud, formed by the
collision of two streams in the warm neutral medium (WNM). The model is
intermediate between a Lagrangian and an Eulerian description, as it
follows the collapse of the cold cloud material as soon as it exceeds
its Jeans mass, but at the same time allows for the addition of fresh
material, coming from the continuing WNM streams (``the inflows''),
through the boundaries of the collapsing region. As the cold gas
collapses and reduces its size, we only add to it the material entering
through its instantaneous, reduced boundaries, while the rest of the
inflow material is assumed to be deposited in an envelope, whose
evolution we ignore. Thus, the model accounts for the fact that a
``cloud'' is {\it not} made of the same material throughout its
evolution, but rather is constantly accreting fresh material from its
environment, as has been proposed by several studies \citep[e.g.,] []
{VS+09, VS+10, Smith+09, Goldbaum+11}. Within this scenario, we follow
the evolution of the material that initially begins to collapse, to which
we will, for convenience, refer to as ``the cloud'', although it must be
borne in mind that the entire system consists of this collapsing region
plus the material added to the envelope during the evolution. Thus, the
entire system does not contract because of the material continuously
added to it. 
}

We assume that the clouds {\bf are born with a density of $n=100~\ppcc$
and a temperature of $\sim40~$K,\footnote{This temperature is obtained
considering the heating and cooling processes by \citet{KI00}.}} 
representative of the cold atomic medium (CNM). In Paper I, the flows
were assumed to continue indefinitely, as is done in many numerical
simulations \citep[e.g.,][]{AH05, Hennebelle+08}.  Instead, here we
assume that the flows subside after 25 Myr, somehow mimicking the
duration of the accretion flow that a parcel in the ISM may be subject
to when traversing a 1-kpc spiral arm at a speed of $\sim 20
\kms$ (the spiral pattern speed with respect to the gas at the Solar
circle). The mass accretion rate onto the cloud is given by 
$\dot{M}_{\rm inf} = 2\rho_{\rm WNM} v_{\rm
inf} (\pi R_{\rm inf}^2)$, where $\rho_{\rm WNM}$ is the WNM density 
($=2.1\times10^{-24}~\gpcc$, which corresponds to  $n = 1~\ppcc$
assuming a mean particle mass of 1.27), 
$v_{\rm inf}$ is the inflow velocity ($=4.5~\kms$, obtained from the
calibration; see below), and $R_{\rm inf}$ is the radius of the
inflow, assumed to have a cylindrical shape. This inflow continues to
feed the cloud for 25 Myr, increasing its mass. We assume that the
cloud begins to undergo global gravitational collapse as soon as it reaches its
Jeans mass which, for a planar cloud, is given by \citep{Larson85}
\beq \label{eq:MJ}
M_{\rm J}=4.67 \frac{c_{\rm s}^4}{G^2 \Sigma},
\eeq
where $c_{\rm s}$ is the sound speed in the cloud, assumed constant and
uniform ($c_{\rm s}(T=40~K)=0.38~\kms$), and $\Sigma = \Mc(t) / 
\pi \Rc^2(t)$ is the surface density, with
$\Mc(t)$ and $\Rc(t)$ being the instantaneous mass and radius of the
cloud, respectively. Note that we deliberately do not consider
turbulent support, as one essential feature of the model is that the
large supersonic velocities that develop in molecular clouds are the
result of the collapse, and thus do not provide support. We also assume
that, because the initial turbulence is transonic (see below), it does
not provide a significant source of additional support. Finally, note
also that, once the cloud has started to collapse, its radius shrinks,
and so we only consider the mass inflow across its instantaneous
cross section, assuming that the rest of the material goes into a
medium-density ($\sim 100 \ppcc$) cloud envelope, which is not included
in the collapse calculation.

%

We furthermore assume that the cold dense gas is turbulent, due to the
combined action of various instabilities \citep{Vishniac94, WF00,
Heitsch+06, VS+06}, with a moderate Mach number $\Mrms =3$. This is
consistent with observations of the velocity 
dispersion in the cold neutral medium \citep{HT03}.  Note that this
Mach number is significantly lower than the typical Mach numbers, $\Mrms
\sim 10$--20, usually associated with molecular clouds, which in
our model correspond to infall velocities rather than to random
turbulence. We stress that numerical simulations in general
\citep{KI02, AH05, VS+06, VS+07, Banerjee+09} show that the turbulent
Mach numbers produced in the atomic precursor of a MC by the flow
collision are substantially smaller than those observed in
MCs. As discussed in
\citet{VS+07}, such high Mach numbers are only observed in cloud
evolution simulations as a consequence of the gravitational
contraction. As a consequence of the turbulence in the dense gas, we
assume that the cloud develops a lognormal density PDF of the form
\beq \label{eq:pdf}
P(s)=\frac{1}{\sqrt{2\pi \sigma_s^2}} ~ {\rm exp} \left[- \frac{(s -
s_{\rm p})^2}{2\sigma_s^2} \right], 
\eeq
where $s \equiv \ln (\rho / \rhoavg)$, $s_{\rm p}= \ln (\rho_{\rm p}
/\rhoavg)=-\sigma_s^2/2$, and $\sigma_s^2= \ln (1+ b^2
\mathcal{M}_{\rm rms}^2)$, with $\rho_{\rm p}$ being the peak
density, $\rhoavg$ the mean density, and $b$ a proportionality
constant related to the compressibility induced by the turbulent forcing
\citep{Federrath+08,Federrath+10}. For simplicity, we consider only
compressible modes, i.e. $b=1$.

As in other recent SFR models \citep{KM05, PN11, HC11, FK12},
we assume that the high-density tail of the PDF is responsible for the
instantaneous SFR in the cloud, which is calculated as
\beq
{\rm SFR}(t) = \frac{M(n > \nsf, t)}{\tff(\nsf)},
\label{eq:SFR}
\eeq
where $\nsf$ is a threshold number density for defining the mass involved in
the instantaneous SFR, and $\tff(\nsf)$ is the free-fall time at number density
$\nsf$. Note that $\nsf$ represents neither 
the mean density of the cloud nor the typical density of the clumps
that form stars. Instead, it is a {\it free parameter} of the model
indicating the density above which the collapse time can be considered to
be negligibly small compared to the evolutionary timescale of the
system. That is, since the cloud
contains a distribution of density fluctuations caused by the
turbulence, the densest among these have the shortest collapse
timescales. The parameter $\nsf$ represents the density
fluctuation level whose collapse time can be considered as
``instantaneous'' in the model. On the other hand, the typical density of a
star-forming cloud or clump is represented by the {\it peak} of the
density PDF, and is generally smaller than $\nsf$, except
at the very last stages of the collapse of a model cloud, when its mean
density reaches very high values (see below).


The threshold number density $\nsf$ was calibrated in Paper I by matching
the evolution of the model to the results of the numerical simulations
--specifically, the evolution of the SFR, and the gaseous and stellar
masses.  The best match was found to occur for $\nsf =10^6 \ppcc$ (for
which the free-fall time is $\tff \approx0.03~\Myr$), and this value was
left fixed thereafter. Here we continue to use that value. 

In addition, in eq.\ (\ref{eq:SFR}), $M(n>\nsf, t)$ is the mass of the
material at densities above $\nsf$,
and given by
\beq
M(n>\nsf, t)= f \Mc(t),
\label{eq:Mhigh}
\eeq
where $f$ is the mass fraction at densities
above $\nsf$, given by\footnote{Note that this equation in Paper I contains a
 typographical error. The form written here is the correct expression.}
\beq
f = \frac{1}{2} \left[ 1-{\rm erf} \left( \frac{
s_{\rm SF}-\sigma_s^2/2}{\sqrt{2} \sigma_s} \right) \right],
\label{eq:f}
\eeq
and $s_{\rm SF} \equiv {\rm ln}(\rho_{\rm SF}/\rhoavg)$ \citep[see
also][]{Elmegreen02, KM05, Dib+11}. 

As mentioned above, in contrast with the models by \citet{KM05}, \citet{PN11}, and
\citet{HC11}, which considered stationary clouds, here we assume that
the cloud is collapsing, keeping in mind that it is a sheet-like object,
and so its collapse proceeds more slowly than that of a
three-dimensional object of the same volume density \citep{BH04,
Toala+12, Pon+12}. In the model, we numerically solve the free-fall
motion of the sheet-like cloud. 

As a consequence of its collapse, the mean density of the cloud
increases with time, causing the density PDF to systematically shift
towards higher densities. Thus, in our model \citep[based on the
notion of hierarchical gravitational collapse;][] {VS+09}, the {\it
global} collapse of the cloud is represented by the fact that the mean
density of a cloud or clump increases over time, while the {\it local}
collapses of the densest regions are represented by the calculation of
the instantaneous SFR, performed by considering the mass above $\nsf$
and dividing it by the free-fall time at this density. This 
treatment implies that the SFR in the model is an increasing
function of time,
since the area under the PDF at densities higher than $\nsf$
increases as the mean density increases.  
The total mass in
stars at time $t$ in the model is thus given by
\beq
\Mstar(t) = \int_0^t {\rm SFR}(t^\prime) dt^\prime.
\label{eq:Mstar}
\eeq
From this stellar mass,
the number of massive stars (with a representative
mass of $17~\Msun$) is computed using a standard IMF 
\citep{Kroupa01}, with lower and upper mass limits 
of 0.01 and $60~\Msun$, respectively. In turn, this allows us
to compute the mass evaporation rate, $\dot{M}_{\rm I}(t)$, by these
stars using the prescription from \citet{Franco+94} for the evaporation
rate induced by a single massive star of age $\hat t$  located near the cloud edge, given by
\beq \label{eq:Midot}
\dot{M}_{\rm I}(\hat{t}) \simeq 2 \pi R^2_{\rm S,0} m_{\rm p} c_{\rm
s,I} \navg 
\Big( 1+\frac{5 c_{\rm s,I} \hat{t}}{2 R_{\rm S,0}} \Big)^{1/5}
\eeq
where $c_{\rm s,I}$ is the sound speed in the ionized gas
($=12.8~\kms$), $\navg$ is the mean number density of the cloud,
$m_{\rm p}$ is the proton mass, and $R_{\rm S,0}$ is the initial
Str$\ddot{\rm o}$mgren radius in a medium of density $\navg$. To
calculate this radius, as in \citet{Franco+94}, we additionally assume a
recombination coefficient for the ionized gas $\alpha_{\rm B}=2.6
\times 10^{-13} \, {\rm cm}^{-3} \, {\rm s}^{-1}$ and a representative
value of the UV Lyman-continuum photon flux $S_*=2 \times 10^{48} \,
{\rm s}^{-1}$, corresponding to our generic massive star. Finally, to
get the total ionized mass we integrate eq.\ (\ref{eq:Midot}) over the
lifetime of each massive star formed and add the contributions from all
active massive stars.

\begin{figure}[!ht] 
\begin{tabular}{c}
\plottwo{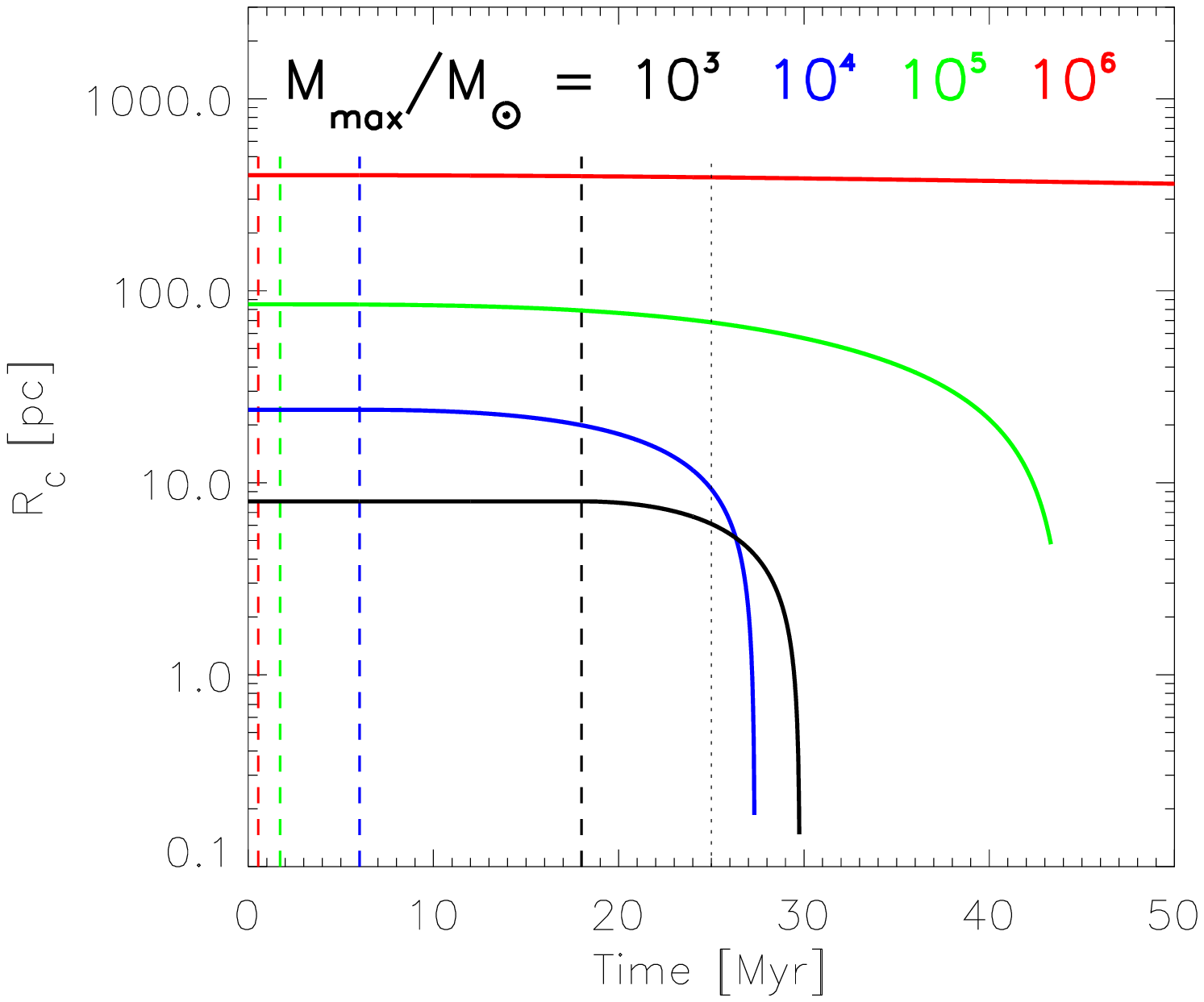}{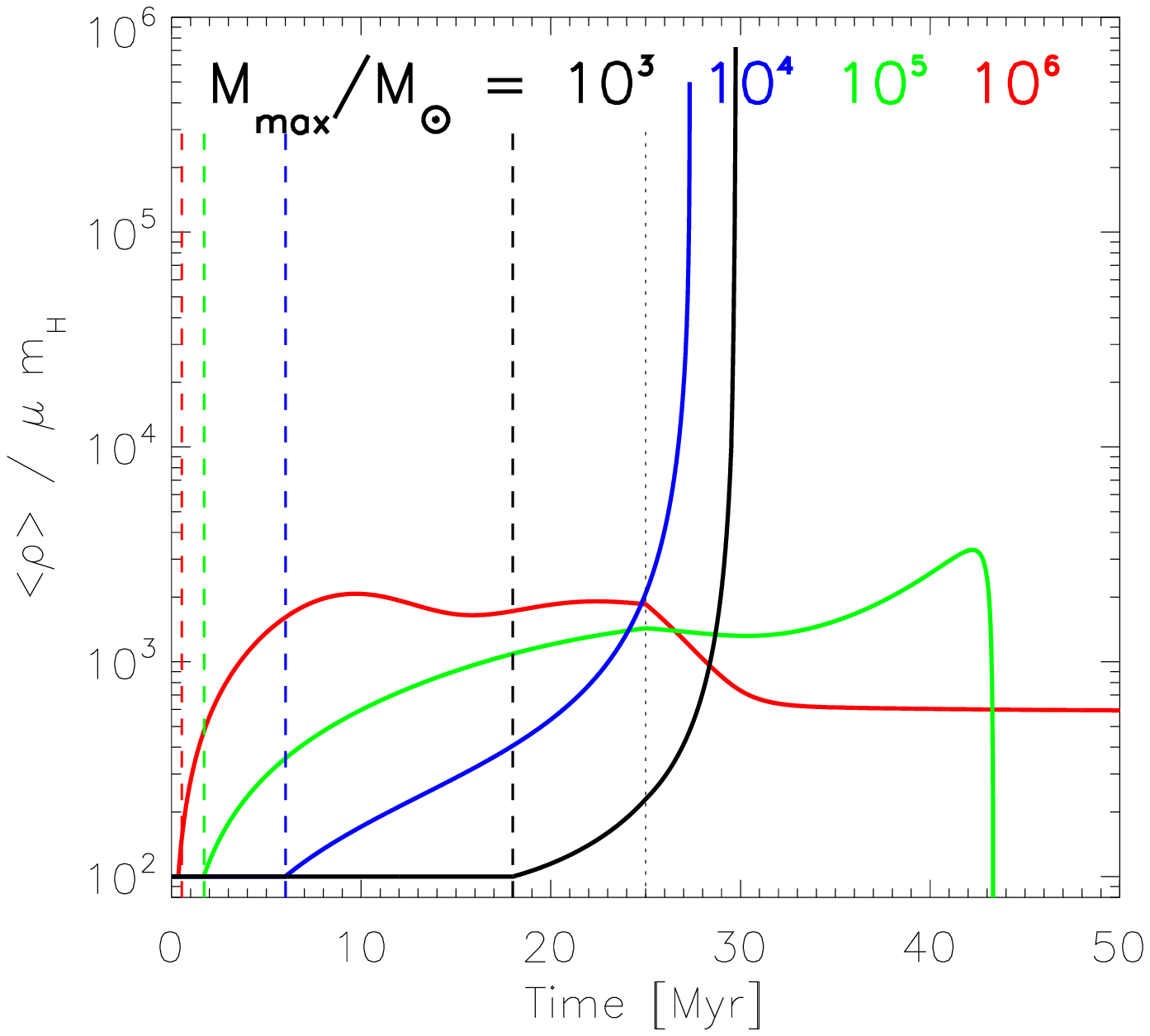}
\end{tabular}
\caption{\label{fig:Rad-Dens} {\it Left:} Time evolution of the radius
for clouds with $\Mmax=10^3$, $10^4$, $10^5$, and $10^6 \, \Msun$
(black, blue, green and red lines respectively). The vertical dashed
lines represent the time at which the cloud reaches its Jeans mass and
begins contracting,
whereas the vertical thin dotted black line 
is the time at which the
accretion stops (at $t=25 \, \Myr$). {\it Right:} Time evolution of the
density. The line colors have the same meaning as in the left panel.}
\end{figure}

With the above ingredients, the instantaneous mass of the cloud is the
result of the competition between addition of fresh gas by accretion,
the consumption by star formation, and the evaporation from the massive
stars. Thus, the cloud mass evolves according to
\beq 
M_{\rm C}(t)=\int_0^t \dot{M}_{\rm inf}(t')~ {\rm d}t' -
\Mstar(t)-M_{\rm I}(t). 
\label{eq:Mc}
\eeq
We numerically integrate eq.\ (\ref{eq:Mc}), together with eqs.
(\ref{eq:Mstar}) and (\ref{eq:Midot}), to obtain the temporal
evolution of a specific cloud, until it is finally
dispersed.

\begin{figure}[!ht]
\begin{centering}
\begin{tabular}{c}
\epsfig{file=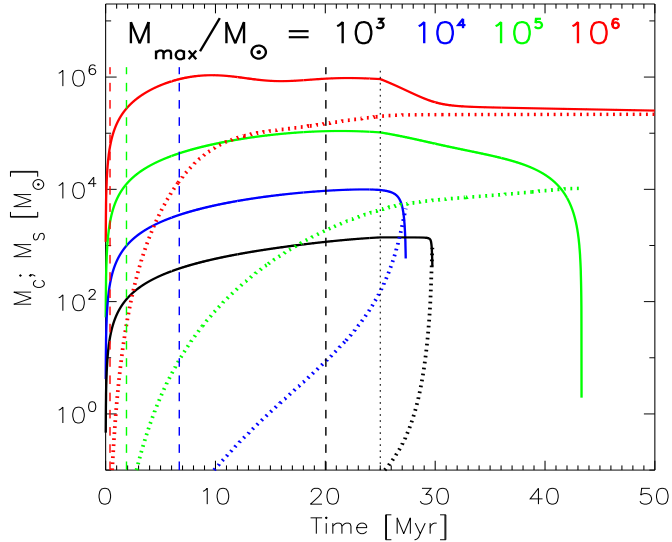,height=8cm,width=10cm}
\end{tabular}
\caption{\label{fig:Mc-Ms} Time evolution of the dense gas mas (solid
lines; eq.\ [\ref{eq:Mc}]), and mass in stars (dotted lines) for clouds
with $\Mmax=10^3$, $10^4$, $10^5$, and $10^6 \, \Msun$ (black, blue,
green and red lines respectively). The vertical dashed
lines represent the time, $\tSF$, at which the cloud starts to form stars,
whereas the vertical thin dotted black line is the time at which the
accretion stops (at $t=25 \, \Myr$).}
\end{centering}
\end{figure}

As emphasized in Paper I, the main controlling parameter of this model
is the total mass involved in the process, which, for fixed values of
$\rho_{\rm WNM}$ and $v_{\rm inf}$ is controlled by the cylinder radius
$\Rinf$. In what follows, we thus choose the required value of
$\Rinf$ to obtain the reported maximum cloud mass, which is the
maximum mass reached by the cold, dense gas during the model's
evolution, and labeled $\Mmax$. Since the numerical integration of the
model takes only a few seconds on a desktop computer, it allows us to
sweep parameter space using hundreds of models of different masses, a
task that cannot be undertaken with full numerical simulations.

Figure \ref{fig:Rad-Dens} shows the evolution of the cloud radius
and mean density for representative clouds
with $\Mmax=10^3$, $10^4$, $10^5$, and $10^6 \, \Msun$. A number of
features are worth noticing. First, from the left panel of Fig.\
\ref{fig:Rad-Dens}, it is seen that the radius of a cloud remains
essentially constant over more than 10 Myr of evolution, during which
the cloud is accreting mass, until it reaches its Jeans mass.
Afterwards, the cloud's radius begins to decrease at an accelerated
pace, with its mean density increasing, as shown by the right panel of
Fig.\ \ref{fig:Rad-Dens}. It is noteworthy that this evolution implies
that {\bf the material constituting an initially medium-sized cloud, of
size $\sim 10$ pc} and mass a few thousand $\Msun$,
such as Perseus or Ophiuchus, should evolve into a
massive-star-forming clump, of density $\sim 10^5$--$10^6 \ppcc$ and
sizes $\lesssim 1$ pc, such as the massive clumps studied by
\citet{Wu+10}, as shown in Fig.\ 7 of Paper I. {\bf It must be borne in
mind, however, that, by the time the cloud has contracted to a massive
clump, it is embedded in the envelope that has been added to the
cloud's surroundings by the WNM inflows.}

Figure \ref{fig:Mc-Ms}, in turn, shows the evolution of the dense gas
mass and the stellar mass for these models.  It is seen that in all but
the most massive model (the one with $\Mmax=10^6 \, \Msun$), the cloud
mass increases until the time when the stellar ionizing feedback begins
to rapidly erode the dense gas mass, causing it to decrease again. The
exception to this behavior is the model with $\Mmax=10^6 \, \Msun$, for
which the dense gas mass stops increasing at $t\sim 10$ Myr. This is due
to the fact that its mass is so large that the mass fraction at high
density in this cloud allows for the formation of massive stars and the
corresponding erosion even before the cloud has had time to contract
significantly\footnote{Recall that in the model, the clouds are assumed
to have sheet-like geometry, and that their collapse is given by the
expression corresponding to such geometry \citep{BH04}, which is slower
than the collapse for a spherical geometry \citep{Toala+12, Pon+12}.}.


In the next section we now discuss the evolution of the SFR and the SFE
for the models as parameterized by their mass.

\section{Model Predictions} \label{sec:predictions}

\subsection{Mass Dependence of the Star Formation Rate} \label{subsec:SFR}

%

In this and the following sections, we consider a collection of models 
of various masses, and focus on the variation of the maximum and
time-averaged values of the SFR and the SFE (cf.\ Sec.\
\ref{subsec:SFE}) as a function of $\Mmax$. The time averages we
consider cover the time span between the formation of $0.01 \Msun$ (the
lower limit in the IMF considered) of stellar products and the
destruction of the cloud (see Fig. \ref{fig:Mc-Ms}). To simplify
the discussion, we will refer to ``low-mass clouds'' as those with
$\Mmax \lesssim 10^4 \, \Msun$; to ``intermediate-mass'' clouds as those
with maximum masses in the range of $10^4-10^5 \, \Msun$, and to
``massive clouds'' as those with $\Mmax \gtrsim 10^5 \, \Msun$.

It is important to note that, so far, we have referred to our
models simply as ``clouds''. However, in this section, in which we try
to predict characteristic values of the SFR and SFE (characterized by
their time averages) in {\it molecular} clouds of different masses, it
is important to define the time interval during which the clouds can be
considered as ``molecular'', so that the time averages are computed over
this interval. Unfortunately, in our one-zone model without chemistry,
there is no direct way to determine this time. Thus, we instead take the
beginning of the averaging interval as the time at which the clouds
begin to form stars, $\tSF$. We have verified that this is a reasonable
proxy for determining when the clouds begin to be sufficiently
molecular by using the density PDF to compute the mass fraction at
densities $n > 10^3 \ppcc$---which can be reasonably assumed to be
already in molecular form---at the time SF starts, finding that in all
models this mass fraction is $\ge 0.2$.
Thus, hereinafter we will refer to
clouds after $\tSF$ as ``star-forming molecular clouds'' (SF-MCs), and
to clouds in previous stages as ``precursor'' clouds.

\begin{figure}[!ht] 
\begin{tabular}{c}
\plottwo{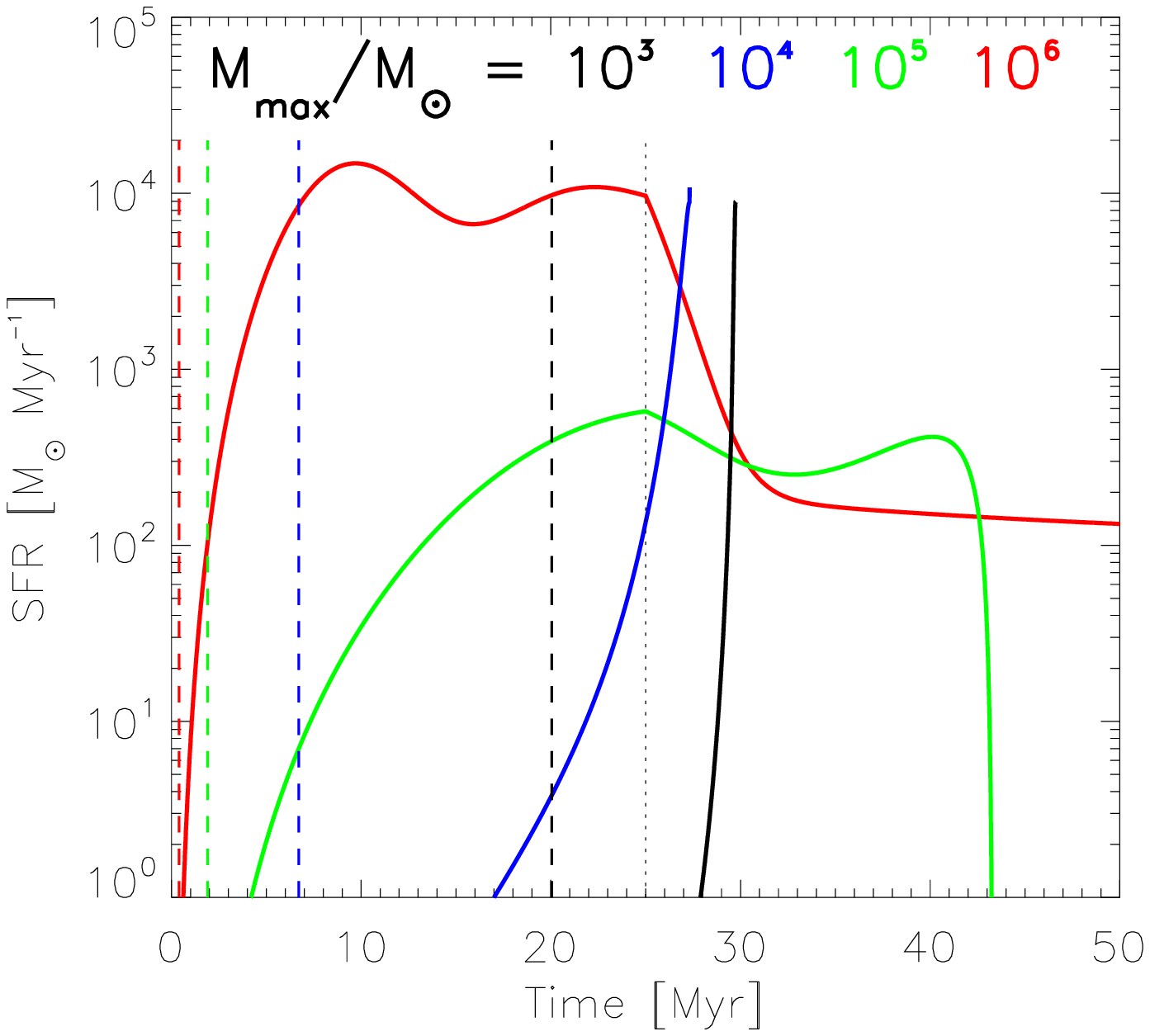}{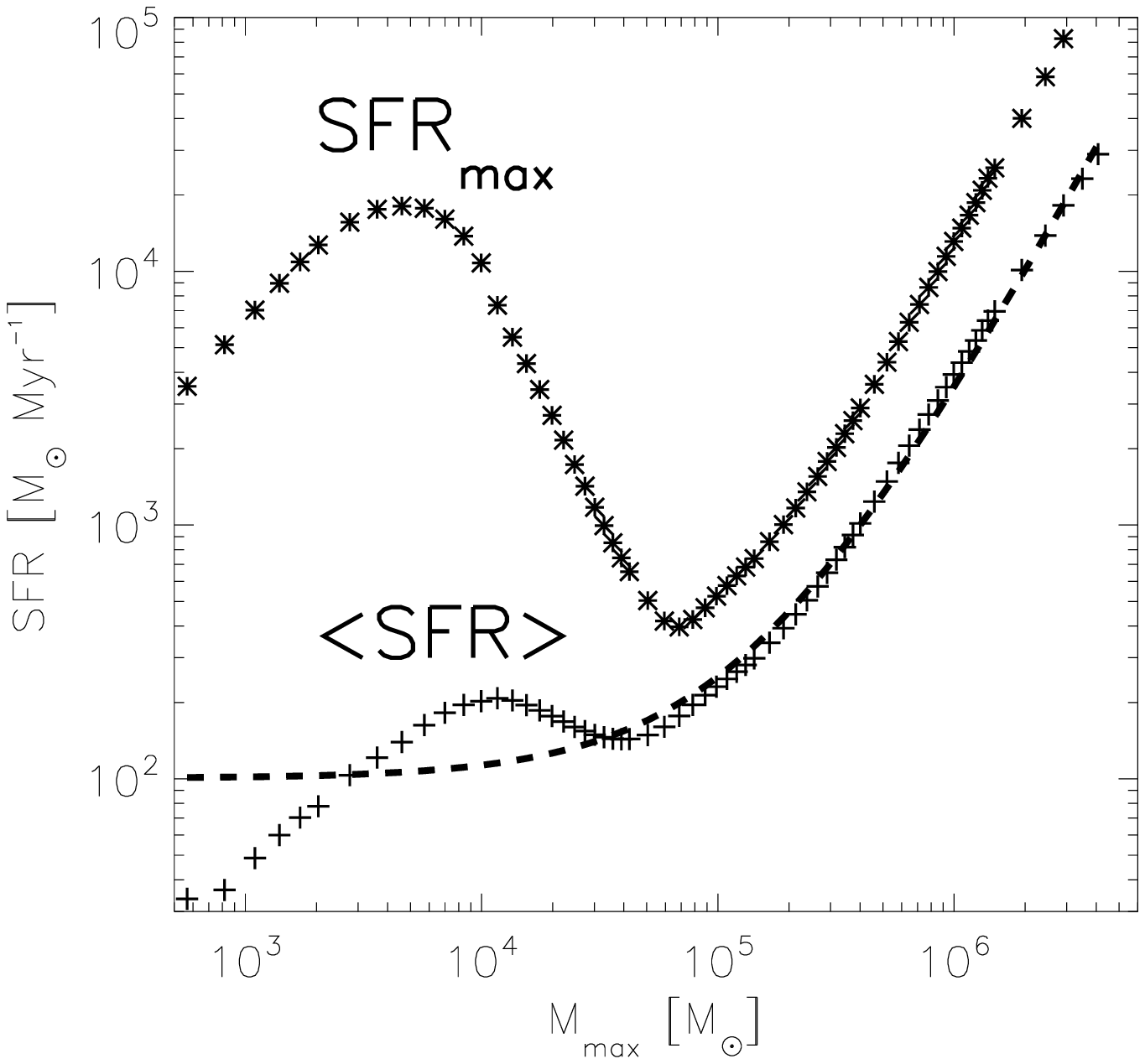}
\end{tabular}
\caption{{\it Left:} Time evolution of the SFR for clouds 
with $\Mmax=10^3$, $10^4$, $10^5$, and $10^6 \, \Msun$ (black, blue,
green and red lines respectively). The vertical dashed lines
represent the time, $\tSF$, at which the cloud starts to form stars,
whereas the verical thin dotted black line is the time at which the
accretion stops (at $t=25 \, \Myr$). {\it Right:} Maximum and
time-averaged SFR asterisks and plus
symbols respectively) as a function of the maximum mass achieved by each
model cloud. The averaging is performed over the period during
which the clouds form stars.}
\label{fig:SFR}
\end{figure}

In the left panel of Fig.\ \ref{fig:SFR}, we show the evolution of the
SFR for clouds of maximum masses $\Mmax = 10^3,~10^4,~10^5$, and
$10^6~\Msun$. We observe that low-mass SF-MCs 
have a very low SFR over
most of their evolution, and end their evolution with a short SF
burst. This can be understood because, due to their low mass, these
clouds can only reach large SFRs when a substantial fraction of their
mass is involved in SF. This can only occur when their mean density has
become comparable to $\nsf$, as shown in the right panel of Figure 
\ref{fig:Rad-Dens}. At this point, these clouds are
quickly destroyed by the first massive stars. This also means that the
consumption of the gas in these clouds is mostly due to SF, rather than
to evaporation by feedback from the massive stars.

On the other hand, intermediate- and high-mass SF-MCs increase their SFR
for the first $\sim 10$ Myr, and then reach a plateau, remaining there
for roughly 10--20 Myr more. This is because, due to their higher
masses, they can have larger SFRs since earlier times (even a small
fraction of their mass corresponds to a large enough mass involved in
active SF).  This implies that massive SF starts earlier in these
clouds. Nevertheless, due to their larger masses and accretion rates,
they are not completely destroyed, although these clouds do lose some 
of their mass when the first massive
stars appear.  Moreover, this partial
mass dispersal causes a decrease in the SFR, and therefore the cloud
evolution reaches an approximately stationary SFR for a significant part of
their lifetimes. Thus, in these clouds, a larger fraction of the dense
gas consumption is due to evaporation, compared to the low-mass clouds.
Note, however, that accretion plays a fundamental role in this behaviour
since, in experiments where we have cut the accretion shortly after the
onset of SF, the massive stars quickly destroy the clouds.

In the right panel of Figure \ref{fig:SFR}, we show the maximum and the
time-averaged values of the SFR as a function of $\Mmax$. Note that each
point in this figure corresponds to a full integration of an individual
model. 
We note that, for low-mass SF-MCs, the maximum SFR 
is much larger than its
average values, because of the short but intense SF burst that
characterizes the end of the evolution of these clouds. Instead, for
more massive SF-MCs, the peak and the average SFR are similar, due to
the prolonged epoch of roughly constant SF that occurs in these clouds.
As a reasonable approximation, the time-averaged SFR can be fit
by a power law of the cloud mass, given by
\beq
\langle {\rm SFR} \rangle \approx 100 ~\Big(1+ \frac{\Mmax}{1.4 \times 10^5 \, \Msun}
\Big)^{1.68} \, \Msun \, \Myr^{-1},
\label{eq:fit_SFR}
\eeq
which is shown as the dashed line in the right panel of Fig.\ \ref{fig:SFR}.

\subsection{Mass Dependence of the Star Formation Efficiency}
\label{subsec:SFE} 

\begin{figure}[!ht] 
\begin{tabular}{c}
\plottwo{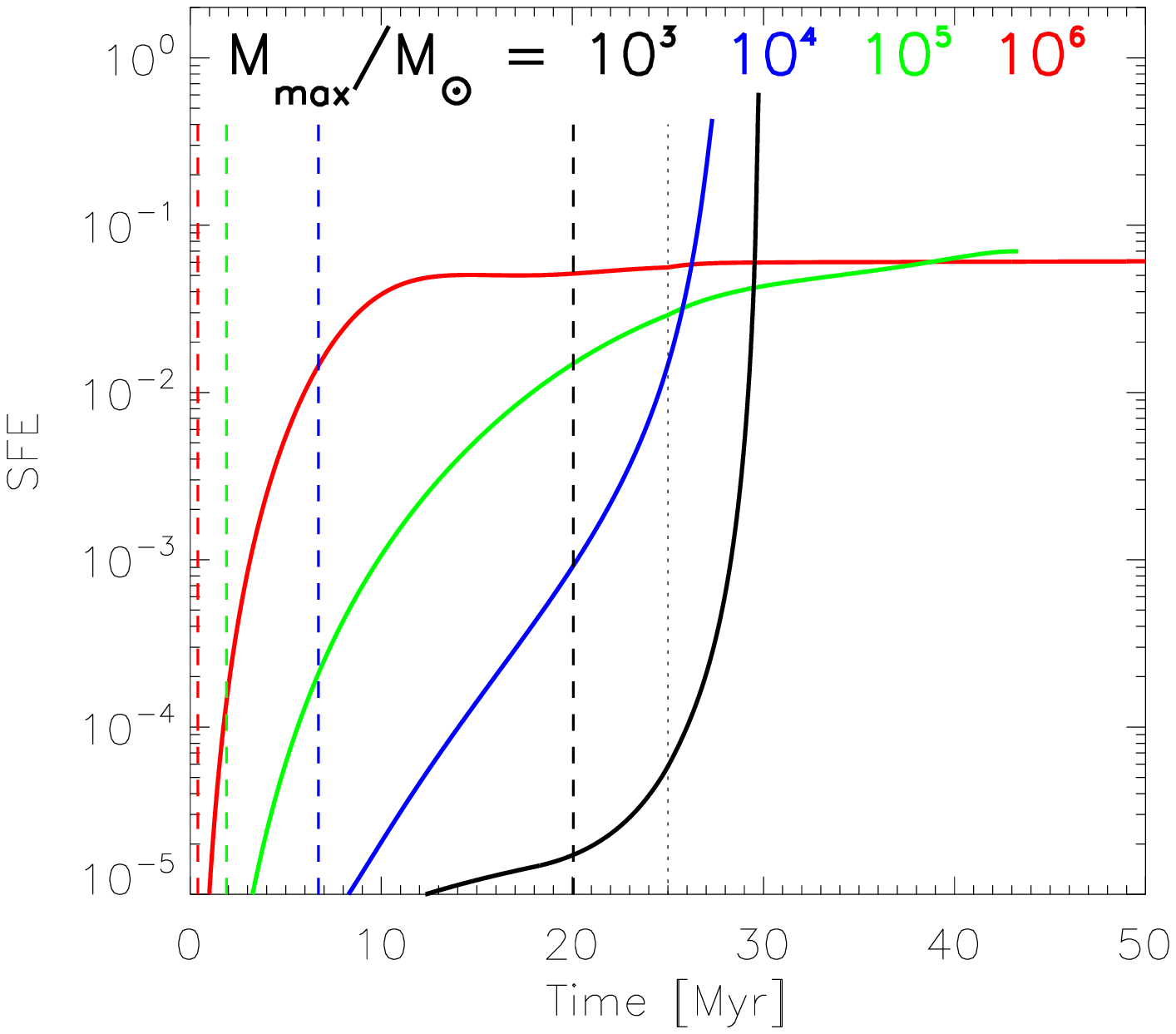}{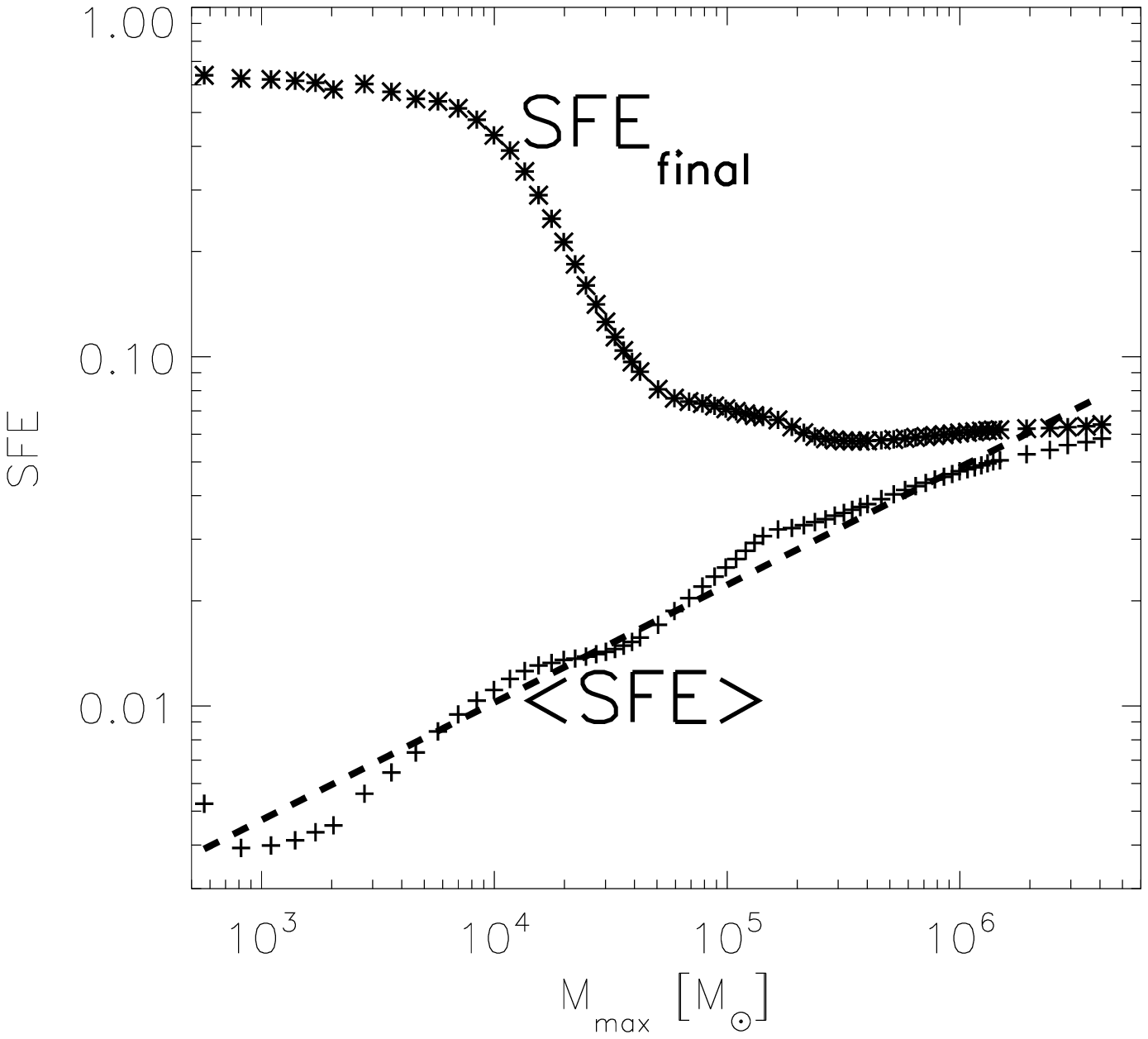}
\end{tabular}
\caption{ {\it Left:} Time evolution of the SFE for clouds with
$\Mmax=10^3$, $10^4$, $10^5$, and $10^6 \, \Msun$. The symbolism is the
same as in Fig.\ \ref{fig:SFR}. {\it Right:}  Maximum and mean
SFEs as a function of the maximum cloud mass.}
\label{fig:SFE}
\end{figure}

We now turn to the mass dependence of the SFE. As in Paper I, we define
the instantaneous SFE as
\beq \label{eq:SFE}
{\rm SFE(t)} = \frac{\Mstar (t)}{\Mc (t) + \Mstar (t) + \Mi (t)},
\eeq
where all the quantities are time-dependent. The left panel of Fig.\
\ref{fig:SFE} shows the time evolution of the SFE for models with
$\Mmax = 10^3,~10^4,~10^5$, and $10^6~\Msun$. From this figure, we see
that in the low-mass SF-MCs, the final star formation 
burst (see right
panel in Fig.\ \ref{fig:SFR}) produces large {\it final} efficiencies
($\lesssim$60\%), although this is not in contradiction with
observations, as it is not possible to observationally determine the SFE
of a cloud/cluster system after the gas has been dispersed. On the other
hand, for the more massive SF-MCs, the SFE reaches a 
saturated value of
$\sim 6\%$. The SFE can saturate due to the interplay between the gas
evaporation by the feedback and the accretion of fresh gas, so that the
masses of the cloud and of the stellar component increase
simultaneously, keeping the instantaneous SFE approximately constant.

The right panel of Fig.\ \ref{fig:SFE} shows the final and time-averaged
efficiencies for the SF-MCs as a function of their 
masses. As in the
right panel of Fig.\ \ref{fig:SFR}, each point in this plot represents
the full temporal integration of one model with a given radius $R_{\rm
inf}$, which reaches the maximum dense gas mass indicated by its
horizontal coordinate. From this figure, we see that, although the {\it
final} instantaneous SFEs of the low mass clouds are much higher than
those of the high-mass ones, the time-averaged values of the SFE
increase monotonically with $\Mmax$. The time-averaged SFE should be
representative of the result of observing a MC ensemble of random
ages, and thus represent the average value of the SFE observed for
MCs of the indicated mass. We see that $\langle {\rm SFE} \rangle$ is
well fit by a power law of the form
\beq
\langle {\rm SFE} \rangle \approx 0.03~\Big( \frac{\Mmax}{2.5 \times 10^5 \, \Msun}
\Big)^{0.33},
\label{eq:fit_SFE}
\eeq
so that we obtain time-averaged SFEs in the range of 0.5-6\% for the
range of maximum cloud masses shown in Fig.\ \ref{fig:SFE}, consistent
with observational determinations for GMCs in general \citep[see, e.g.,
the compilation by][]{FK13}. Thus, our model predicts that the
time-averaged SFE should increase with the cloud mass, albeit slowly,
with a scatter that might correspond not only to observational errors,
but also to the variation of the SFE over the evolution of the clouds.

The exponent in eq.\ (\ref{eq:fit_SFE}) is close to the
value predicted analytically by \citet{Fall+10} of 0.25 for the case of
feedback dominated by ionization heating. The difference may be due to
the fact that those authors considered stationary energy (or momentum,
for the case of momentum-dominated feedback) balance, while here we take
the additional step of considering the time evolution of the
feedback, and/or to the different assumed geometries (flat in our case).

\subsection{Star formation rate-dense gas mass correlation}
\label{subsec:Gao}

In Paper I we showed that the evolution of our model clouds with maximum
dense gas masses $\Mmax \sim 2000~\Msun$, when plotted in the
Kennicutt-Schmidt (KS) diagram, took them from the locus of clouds
forming low-mass stars, such as Perseus, Lupus, Serpens, and Ophiuchus
\citep{Evans+09}, to that of clouds forming massive stars, such as the
Orion A cloud and the clumps investigated by \citet{Heiderman+10}.  We
now investigate whether, collectively, our clouds conform to observed
star formation correlations found for averages over large volumes. This
is important because those correlations are often interpreted as the
result of a sustained low value of the SFE due to global turbulent
support of the clouds \citep[e.g.,][]{Krumholz+09, Krumholz+12}, while
our model clouds are in global collapse, and their SFR and SFE are not
constant, but rather, time-dependent. In this case, one can ask whether
the {\it time-averaged} (over their star-forming epoch) SFR and SFE of
our model clouds are consistent with the observed SF correlations.

\begin{figure}[!ht] 
\begin{centering}
\begin{tabular}{c}
\epsfig{file=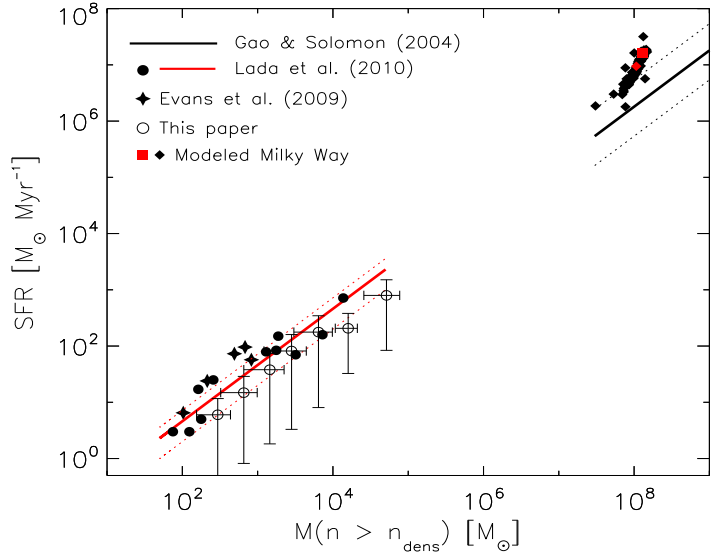,height=8cm,width=10cm}
\end{tabular}
\caption{\label{fig:SFR-Mdens} 
 SFR as a function of dense gas mass ($\Mdens=M(n \geq \ndens)$, with
$\ndens = 10^4 \, \ppcc$) for low- to intermediate-mass model clouds
(open circles with error bars; see text). In the lower-left corner,
corresponding to individual cloud masses, the filled stars correspond to
data from \citet{Evans+09}, while the filled circles correspond to the
cloud sample studied by LLA10, with the red solid line denoting the mean
fit reported by those authors, and its dispersion represented by the red
dotted-lines. In the upper-right corner, corresponding to galactic
masses, the solid black line shows the scaling found by GS04 (eq.\
[\ref{eq:Gao}], with $\ndens = 3 \times 10^4 \, \ppcc$), the black
dotted-lines showing the scatter of their observational sample. The
filled red square shows the position of the modeled Milky Way given by
eq.\ (\ref{eq:MW_SF}). Finally, the filled black dimonds represent Monte
Carlo realizations of cloud ensembles taken at random evolutionary
stages (in the SF-MC stage), with the red filled diamond giving the
average value of these experiments.}


\end{centering}
\end{figure}

One important such correlation is the one found by \citet[][herafter
GS04]{Gao+04} who, in a sample of luminous and ultraluminous infrared
galaxies, as well as of normal spirals, found a
linear relationship between the IR luminosity (a tracer of the SFR) and
the HCN luminosity (a tracer of the dense [$n \ge \ndens = 3 \times
10^4~\ppcc$] gas mass), implying that
\beq
\SFR \approx 180~\left(\frac{\Mdens}{10^4\, \Msun} \right) \, \Msun \,
\Myr^{-1}, 
\label{eq:Gao}
\eeq
which is a linear relationship between the SFR and $\Mdens$, 
the mass at density $\geq \ndens$ (Fig.\ \ref{fig:SFR-Mdens}).
On MC scales, \citet[] [hereafter LLA10] {Lada+10} found a 
similar linear relationship
(SFR $\propto \Mdens$) for a sample of nearby MCs (see
Fig. \ref{fig:SFR-Mdens}), measuring the gas mass at densities above $n
\ge \ndens = 10^4~\ppcc$ from extinction maps, and estimating the SFR by
counting young stellar objects (YSOs) and dividing by a typical age
spread, $\Delta t \sim 2$ Myr. These authors also found that the SFEs of
their cloud sample fall in the range of 0.1-4\%. 

We wish to compare our model's predictions to these results. To do so,
at the individual cloud level, we attempt to replicate the procedure of
LLA10. Note that, according to our model, both the SFR and the SFE of a
cloud increase over time, and thus the range of efficiencies observed in
LLA10's sample is interpreted as a spread in evolutionary stages. In
turn, this implies that the corresponding SFRs should also correspond to
a range of evolutionary stages. Thus, for each model cloud, we should
consider the range of SFRs that it may have over this range of
evolutionary stages for comparison with the observations. However, we
have the problem that the evolutionary stage of the clouds considered
by LLA10 is not known. To circumvent this problem, we make use of the
fact that our model predicts {\it both} the SFR and the SFE of the
clouds as a function of time. These quantities are different from each
other, because the SFR is a truly instantaneous quantity, while the SFE
involves the integral of both the stellar and gaseous mass accretion
rates (cf.\ eq.\ [\ref{eq:SFE}]). Thus, we can use the instantaneous SFE
of a cloud as a proxy for its evolutionary stage, and then compute the
corresponding SFR predicted by the model at that stage, to compare with
the observationally-inferred SFRs. 

Note, in addition, that the SFR estimates by LLA10 are not strictly
instantaneous values, but rather, averages over the age spread $\Delta
t$. We thus also estimate the SFR not through the instantaneous SFR
predicted by the model, but as the number of stars of age $< 2$ Myr,
divided by this time interval, and compute this estimate at two
different times, one, labeled $t_1$, when the clouds' SFE is 0.1\% and
the other, labeled $t_2$, when the clouds' SFE is 4\%, consistent with
the range of SFEs exhibited by the LLA10 sample. Thus, for each model
cloud we report a range of SFRs. Similarly, the instantaneous mass of
the model clouds varies between these two times, and therefore we also
report a range of masses for each model cloud. We do this for model
clouds with maximum masses in the same mass range --- recall the model
clouds are labeled by the maximum mass they reach during their
evolution --- as the clouds in their sample.
%

%

In Fig.\ \ref{fig:SFR-Mdens} we show, in the lower-left corner, the
range of SFRs computed as described above {\it versus} the range of
dense gas mass (i.e., at densities $n > \ndens=10^4 \, \ppcc$) for our
model clouds (open circles with error bars) and for the cloud sample
studied by LLA10 (filled black circles) and by \cite{Evans+09} (filled
black stars).\footnote{LLA10 report the dense gas mass, and we take the
number directly from them. However, \cite{Evans+09} only report total
masses, and hence we estimate $\Mdens$ for their cloud sample using the
same procedure as for our model clouds; that is, we assume a lognormal
PDF centered at the mean density of the clouds reported by those
authors, and with width corresponding to a Mach number of 3.} The ranges
of SFR and $\Mdens$ between $t_1$ and $t_2$ for the model clouds are
indicated by the error bars, while the open circles are the
mean of these quantities between these two times. It is seen that the
slope of the ensemble of models is nearly unity, similarly to the fit by
LLA10, and that the locus of the means falls within the scatter reported
by those authors.

As a further test of the model, we can estimate
the {\it total} Galactic SFR it predicts by convolving the
time-averaged SFR for each cloud mass, $\big\langle
\hbox{SFR}\big\rangle\, (M)$, with a suitable cloud mass spectrum. 
This can be then compared with the average relation derived by GS04 for
external galaxies. 

We use the Galactic cloud mass spectrum derived by \citet{Williams+97},
given by
\beq
d \NC = N_0 \left( \frac{M_{\rm U}}{M}\right)^\alpha \, d (\ln M)
\label{eq:W97}
\eeq
where $d \NC$ is the number of MCs with masses in the range $M$ 
to $M+{\rm d}M$, $N_0=63$, $\alpha=0.6$ and $M_{\rm U}=10^6~\Msun$ is the 
maximum assumed mass of GMCs in the Galaxy \citep[see also ][]{MO07}.
The total Galactic SFR is then given by
\beq
{\rm SFR}_{\rm tot} = \int_0^{M_{\rm U}} \langle{\rm SFR}\rangle (M) d \NC 
\label{eq:MW_SF}
\eeq
This exercise gives a global Galactic SFR of $14 ~ \Msun~
{\rm yr}^{-1}$, within a factor of 5 from recent observational
estimates \citep[e.g.,][]{ChP11}. Also, we can compute the total dense
mass above $\ndens$ as
\beq
M_{\rm tot}(n>\ndens) = \int_0^{M_{\rm U}} \langle{\rm M(n>\ndens)}\rangle (M) d \NC
\label{eq:MW_dense_mass}
\eeq
where $M(n>\ndens)$ is given by eqs.\ (\ref{eq:Mhigh}) and (\ref{eq:f}), 
replacing $\nsf$ by $\ndens$.
We find $M_{\rm tot}(n>\ndens) = 1.3 \times 10^8 ~ \Msun$.
The red filled square near the upper-right corner of
Fig. \ref{fig:SFR-Mdens} shows the resulting ``model galaxy'', based on
our model's predicted $\langle$SFR$\rangle$, which is seen to be larger
than the mean scaling by GS04 by a factor $\sim 5$. The discrepancy
is probably due to the strong SF bursts predicted by our model for low-mass
clouds, in which $\sim 40\%$ of the total SFR takes place, according
to the mass spectrum.

A perhaps more precise comparison is provided by a Monte Carlo
integration, taking the SFR and the corresponding mass at random times
for each SF-MC, and integrating again according to the mass spectrum. In
the upper-right corner of Fig. \ref{fig:SFR-Mdens} (filled black
diamonds) we also show a hundred of these experiments, obtaining average
values of $9 ~ \Msun~ {\rm yr}^{-1}$ and $\Mdens = 1.1 \times 10^8 ~
\Msun$ for the total SFR and dense mass gas (filled red diamond),
respectively. Roughly a third of the points generated in this way are
seen to fall within the uncertainties of the GS04
relation. Nevertheless, the set of points always falls above the mean
GS04 relation, and so the model does seem to overestimate the Galactic 
SFR by a factor of 3--5. We discuss this further in
Sec.\ \ref{subsec:limitations}. 

\section{Discussion and limitations} \label{sec:discussion}

\subsection{Implications and insights} \label{subsec:implications}

One important prediction of our model is that GMCs evolve in
such a way that a moderate-size MC eventually
becomes a dense, massive-star-forming clump,
over the 
course of $\sim10$ Myr. We now discuss how this process fits into
our established knowledge about MCs.

\subsubsection{Large- and small-scale collapse}
\label{sec:large_small_coll} 

 The first point to emphasize is that the model is designed to
account for both the collapse of small-scale clumps and cores and the
collapse of the cloud as a whole. This is in line with the notion
advanced by \citet{VS+09} that gravitational collapse in MCs is {\it
hierarchical} so that small-scale, local collapses occur within the
environment of a cloud that is also undergoing collapse as a whole (a
large-scale collapse). The small-scale collapse is described in the
``standard'' way \citep[e.g.,] []{KM05, HC11, PN11}, by assuming that
the cloud contains a distribution of density fluctuations, the densest
of which are undergoing instantaneous collapse, and therefore being
responsible for the instantaneous SFR of the cloud. This amounts for
what is normally referred to as the fragmentation of the cloud.

The large-scale collapse, on the other hand, is accounted for by
directly computing the contraction rate of the whole cloud based on its
average density and corresponding free-fall time. This contrasts with
the models cited above, which assume roughly stationary conditions in
the clouds, and therefore cannot account for any evolutionary features
of the clouds. In our model, the whole evolutionary nature of the
process derives from the fact that the cloud is contracting as a whole.

\subsubsection{Velocity gradients: rotation or infall?} \label{sec:rot_inf}

A widespread notion is that MCs rotate, and that such rotation would
prevent their contraction to clump scales. Indeed, velocity gradients
are ubiquitously observed in MCs \citep[e.g.,] [] {PG97, Rosolowsky+03,
Brunt03, Brunt+09} as well as in dense cores within them \citep[e.g.,] [
see also the review by Belloche, 2013] {Goodman+93, Kirk+10}. However,
as stated above, {\it some} form of contraction must occur in order to
form a massive, dense clump.

Although the velocity gradients are almost always interpreted as
rotation, it is important to remark that there is no {\it a priori}
reason to do this, and in fact, \citet{Brunt03} points out that the
Principal Component Analysis of the velocity structure in MCs is
inconsistent with the signature of rotation in model clouds. On the
other hand, \citet{VS+08} showed that overdense regions in simulations
of driven isothermal turbulence exhibit on average a negative velocity
divergence (i.e., a velocity convergence) whose magnitude is within the
range of velocity gradients reported by \citet{Goodman+93} in cores of
similar sizes, suggesting again that a significant component of the
observed gradients may actually consist of inflow motions, rather than
rotation. Similar conclusions were obtained by \citet{Csengeri+11} for
massive dense cores in Cygnus-X, where they concluded that infall as
well as rotation may be present.

Finally, it should also be noticed that, in general, these infalling
motions are not expected to be spherically symmetrical, as MCs are
observed to consist mostly of filamentary structures \citep[e.g.,] []
{Myers09, Andre+10, Molinari+10, Kirk+13}, and therefore the classical
infall signature \citep[e.g.,] [sec.\ 4.7] {Evans99} should not be
expected in molecular line observations of objects at these scales, so
the failure to detect them does not rule out the possibility that the
velocity gradients observed across clouds correspond in fact to
collapsing motions.

\subsubsection{Delayed and extended star-formation activity}
\label{sec:delayed_SF} 

Another prediction from the model is that the evolution of the clouds
includes a period of time (see left panel of Fig.\
\ref{fig:SFR}) with no significant SF (i.e.,
as {\it precursor} clouds) and, once star formation starts, they are expected
to spend several more Myr at
low SFRs. Interestingly, the low-mass
clouds spend longer times in these states than the high-mass ones. For
example, it can be seen in Fig.\ \ref{fig:SFR} that a cloud with  
$\Mmax = 10^3 \Msun$ takes $\sim 20 \, \Myr$
to start forming
stars, and after that, it spends $\sim 8$ Myr with very low
SFR. This is because the low-mass clouds need to contract by a large
factor to reach a large enough mean density that the mass above the
threshold density $\nsf$ causes a significant SFR. Instead, the
high-mass clouds begin to do so at earlier times, when their mean
density is still relatively low, because even a small fraction of their
mass above $\nsf$ involves sufficient mass for the SFR to already be
significant.

However, this prediction might appear contradictory with the
notion that MCs form stars rapidly after their formation
\citep{HBB01}.  What must be borne in mind here is that the model
follows the evolution of the clouds from their earliest, cold-atomic
stage, which is effectively the precursor of a GMC \citep{VS+06}. Clouds
formed by colliding WNM streams are expected to build up their molecular
content over relatively long timescales \citep[e.g.,] [] {FC86, HBB01,
HH08, Micic+13} and should only become mostly molecular by the time when
they have become strongly gravitationally bound. In the context of our
model, then, the zero- or low-SFR epochs correspond to times when the cloud is
still atomic-dominated, consisting of a collection of moderate-mass,
slowly-star-forming molecular clumps, immersed in an atomic
substrate. The high-SFR stages occur when the cloud is already in a
mostly molecular state, in agreement with the notion that MCs form stars
rapidly, within a few Myr from their formation. The subtle additional
consideration is that the clouds have much longer time spans, but during
most of that time, they are mostly non-molecular, and they are forming
stars at very low rates. We estimate the molecular fraction of the
clouds' mass as a function of time for the sample clouds of masses $10^3$ 
to $10^6 \, \Msun$, and in all cases, by the time SF starts, 20\% or more 
of the mass is ``molecular'' (gas with number density greater than $10^3 \, 
\ppcc$).

%

We emphasize that this result is consistent with the fact that embedded
clusters generally contain a small fraction of older stars, although the
majority of their stars is young \citep{PS99, PS00}. Indeed, in Paper I
we showed that the model correctly recovers the typical age
distributions found in those clusters.  Also in this regard, it should
be noted that the above discussion implies that the gravitational
contraction is likely to start in the mostly-atomic stage. This is
consistent with numerical simulations of cloud formation that also
follow the atomic and molecular fractions, which suggest that molecules
actually form {\it during} the gravitational contraction \citep{HH08},
and with the result by
\citet{GC12} that molecules are in fact not necessary for producing the
cooling needed for gravitational collapse.

Finally, an important point is worth clarifying. The prediction that
massive GMCs, with $10^5$ -- $10^6 \Msun$, have extended periods of star
formation lasting 20 Myr or more may seem to be in conflict with
observations of nearby clouds like Orion and ``dispersed'' regions, such
as Sco-Cen, as it is generally accepted that cloud dispersal by the
recently-formed stars occurs rapidly, within a few Myr. However, this
quick dispersal refers essentially to the immediate gaseous environs
of a newly formed cluster only, as it is also known that SF occurs only
over a small fraction of a cloud's volume \citep[e.g.,][]{Kirk+06},
while the destruction of a large GMC may easily take over 10 Myr. For
example, in the Sco-Cen region, the three main subgroups, Upper
Scorpius, Upper Centaurus-Lupus, and Lower centaurus-Crux, have ages
that differ by more than 10 Myr, and it has been suggested that the
latest events might have been triggered by the earlier ones
\citep[e.g.,][]{PM08}. Thus, SF in the parent GMC might have been going
on for at least that amount of time, suggesting that the lifetime
predicted by our model for massive clouds is reasonable. In the case of
this complex, however, the dispersal has probably been completed
already, as the clusters there are already devoid of gas. But the SF
episode must have lasted at least the length of time spanned by the age
difference between the clusters.

On the other hand, \citet{Kawamura+09} have suggested that GMCs with
masses $\sim 10^5$--$10^6 \Msun$ in the LMC may have lifetimes of $\sim
25$ Myr, with three well-defined stages in terms of their SF
activity. In Paper I we showed that our model, for clouds of those
masses, matches within a factor of 2 the duration of the individual
stages, while the left panel of Fig.\ \ref{fig:SFR} here shows that
indeed this is the time span of the star-forming stages of such clouds.

Thus, we conclude that the cloud evolution predicted by the model is in
very good agreement with many known properties and evolutionary features
of MCs.

\subsection{Limitations} \label{subsec:limitations}

Our SFR model is clearly an extreme idealization of the actual process
occurring in MCs, as it only considers the effects of self-gravity and
photoinonization on the evolution of the clouds. In particular, it
neglects any support from magnetic fields, which are known to retard the
gravitational collapse in comparison with the non-magnetic case
\citep[e.g.,][] {Ostriker+99}, the momentum injection by the
ionizing stellar feedback and by stellar outflows, and the
additional feedback from supernovae and radiation pressure from the most
massive stars. Since all of these processes tend to either counteract
the collapse or to destroy the clouds more rapidly, it is clear that the
SFR and SFE predicted by the model are upper limits to those in real
clouds.

Nevertheless, it is all the more interesting that, within these
important limitations, our model in general predicts values of the SFR
and the SFE, as well as evolutionary features of the clouds (Paper I),
that generally agree well with the corresponding observational
measurements, with the largest deviations occurring when the
time-averaged values of the SFR for all cloud masses are added to
construct a Galaxy-wide SFR. Better agreement is obtained for the
Galaxy-wide SFR when a set of Monte-Carlo realizations is considered,
using values of the instantaneous SFR at random times for each cloud
mass. This suggests that the final SFR burst predicted by the model for
low-mass clouds may be overestimated, and indeed, the time-averaged SFR
predicted for these clouds exhibits a bump at low-to-intermediate cloud
masses (see Fig.\ \ref{fig:SFR}). This suggests that, especially for
these clouds, the effects of magnetic fields and outflows
may be most important. Nevertheless, the general
better-than-order-of-magnitude agreement of the model with the
observations suggests that self-gravity and
photoionizing radiation, the processes considered by the model, are
among the dominant processes controlling the evolution of the clouds and
their star formation activity, with the other processes providing
second-order corrections only.

On the other hand, possibly the most questionable ingredient of our
model is the assumption that the density PDF remains lognormal during
the entire evolution of the clouds, an assumption that appears in
conflict with the well-known result, from both observations and
numerical simulations, that star-forming clouds develop a power-law
high-density tail in their column density distributions
\citep[e.g.,][]{Kainulainen+09, Kritsuk+11, BP+11,
Girichidis+14}. However, in Paper I we argued that turbulence {\it
alone} produces a lognormal, which is the {\it seed} of subsequent
gravitational collapse, and that the power-law tail is a {\it result} of
this contraction. Thus, the mass in this regime perhaps should not be
counted as a seed for subsequent collapse, since it is already
undergoing collapse. In any case, to minimize the impact of this
assumption, in Paper I we calibrated the value of the $\nsf$ by matching
the predictions of the model to the output of the self-consistent
numerical simulations of \citet{VS+10}. The {\it a-posteriori}
confirmation of this procedure is that, using the lognormal, the model
was able to match a variety of observations.

Another important point to recall is that the model assumes clouds with
a flattened geometry, for which the collapse timescale is significantly
longer  than for a spherical geometry, typically by factors of half to 
one order of magnitude, than the
standard free-fall time \citep{Toala+12, Pon+12}. However, this is
probably a reasonable assumption, since most clouds are known to consist
of flattened or filamentary structures \citep[e.g.,][]{Bally+89}. This
suggests that 
another important factor determining the SFR is the non-spherical
geometry of MCs.


\section{Summary and Conclusions} \label{sec:sum}

In this paper, we have presented the predictions for the dependence of
the time-averaged SFR and SFE on the mass of the parent cloud from our
semi-analytical model for the evolution of these quantities in
gravitationally collapsing clouds, introduced in
Paper I. The model assumes that the cloud forms by the collision of two
streams in the WNM, which induces a transition to the cold phase,
forming a cold cloud that becomes turbulent due to various instabilities
\citep{Heitsch+05, Heitsch+06, VS+06}. Soon this turbulent cloud begins
to undergo global gravitational collapse. The collapse is {\it hierarchical},
because the turbulence in the cloud produces density fluctuations that
have shorter free-fall times than the cloud as a whole, and then form
stars before the collapse of the largest scales is completed. The
fraction of the cloud's mass involved in instantaneous SF is determined
by assuming that the density PDF in the cloud is lognormal, and that
only the mass above a certain critical density, $\nsf$, is
instantaneously forming stars. As the cloud collapses, its mean
density increases, so that the PDF shifts to higher densities, causing
the instantaneous SFR to systematically increase in time.

The total amount of gas converted into
stars is distributed among stellar masses accorging to a standard
IMF. The most massive stars produce ionizing radiation, which evaporates
parts of the cloud through HII regions. While all this is happening, the
cloud continues to accrete material from the converging flows. Thus, the
evolution of the cloud is regulated by the competition between addition
of fresh material by the accretion and the gas consumption by the SF
itself, as well as by the evaporation by the ionizing radiation from the
massive stars. The model neglects the magnetic field and any injection
of momentum by the stellar feedback.
 
In Paper I, it was found that the total mass involved in the process is
 the main free parameter controlling the evolution of the clouds and their
SFR. We quantify this parameter by the maximum dense gas mass reached by
the clouds, $\Mmax(n \ge 100 \, \ppcc)$. In the present contribution, we have considered the
evolution, the final values, and the time averages over the star-forming
epochs of the model clouds, of the SFR and SFE predicted by the model,
as a function of the maximum dense gas mass attained by the model
clouds. We have found that low-to-intermediate-mass model clouds ($M
\lesssim 10^4 \, 
\Msun$) spend their early and intermediate evolutionary stages forming 
stars at low rates, while a strong
star formation burst is produced during their final, dense stages
(when they appear as a massive clump within a larger cloud), at which
time massive stars appear 
and quickly destroy the cloud. Therefore, these clouds have a low
time-averaged SFR ($\langle {\rm SFR} \rangle$) but a high final
SFR. Instead, in massive clouds ($M \gtrsim 10^5 \, \Msun$), massive
stars appear from early in their evolution, and thus the ionizing
feedback regulates the SFR almost from the beginning. This leads to a
final SFR comparable with the average. We provided fits to the mass
dependence of the time-averaged SFR and SFE, given by eqs.\
(\ref{eq:fit_SFR}) and (\ref{eq:fit_SFE}).

We then proceeded to investigate the relation between the SFR and
the dense cloud mass, $\Mdens$, for our model clouds, mimicking the
procedure followed by LLA10. These authors estimated the SFR as the mass
in YSOs (in our case, stars younger than 2 Myr) divided by this
time interval. Since the clouds studied by LLA10 span a wide
range in SFEs, we considered our model clouds in the time interval
during which they span the same SFE range.
We found that the mean values of the SFR and the clouds' mass
during this time interval
fall within the error bars of the mean relation reported by LLA10.

We also estimated the total Galactic SFR predicted by our model,
by convolving the SFR (in average or taken 
it at a random time after the onset of star formation) for each cloud
mass with the Galactic cloud mass spectrum by \citet{Williams+97}. The
average of a hundred of these random realiztions is within half an order
of magnitude from the observed Galactic SFR, and  from the scaling
relation found by GS04 for the global SFR vs.\ galaxy mass of a sample
of external galaxies.

With respect to the SFE, we find that for low-mass clouds, in the final
star formation burst, the efficiency reaches final values $\sim
60\%$, although these values are not in conflict with observations
because they correspond to the stage when no dense gas mass is left
around a cluster, at which point it is almost impossible to
observationally know the initial amount of gas mass that went into the
formation of the cluster. The time-averaged SFE, on the other hand, is
$\sim$1\%, consistent with observational determinations performed
on clusters still embedded in their parent clouds
\citep[e.g.,][]{Evans+09}. For massive clouds, the SFE reaches values up
to 6\% (but with averages $\lesssim$5\%), consistent with the upper limits 
of SFE ($ \sim 10\%$) determined in \citet{FK13}.

We next discussed several implications of the model in the context
of well established notions about MCs and their SF activity, arguing
that, although some of the model predictions and implications may seem
to be in conflict with those notions, upon closer examination no
conflict exists, and instead the model offers a new insight about the
evolution of MCs.

As pointed out in Sec.\ \ref{sec:discussion}, the fact that our
extremely idealized model, in which only self-gravity and ionizing
feedback control the evolution of the SFR in the clouds, fits the
observations typically within factors of a few, suggests that
these may be the dominant controlling processes, with other processes,
such as magnetic support and momentum injection from massive-star winds,
probably providing mainly second-order adjustments.

\acknowledgements
We are thankful to the anonymous referee, whose constructive report
helped us in improving the clarity and reach of the predictions and
implications of our model. We also thankfully acknowledge partial
finantial support by UNAM grant IN102912 to M.Z.-A and CONACYT grant
102488 to E.V.-S.

\clearpage

\end{document}